\documentclass[11pt]{article}
\usepackage{amssymb}
\usepackage{amsthm}
\usepackage{amsmath} 
\usepackage{amssymb}
\usepackage{amsthm}     
\usepackage{array}
\usepackage{bbm}
\usepackage{epsfig}
\usepackage{latexsym}
\usepackage{dsfont}
\usepackage{wasysym}
\usepackage{flafter}
\usepackage{accents}
\usepackage{color}
\usepackage{graphicx}
\usepackage{geometry}
\newcommand{\beqan}{\begin{eqnarray*}}
\newcommand{\eeqan}{\end{eqnarray*}}   
\newcommand{\ba}{\begin{array}}
\newcommand{\ea}{\end{array}}
\newcommand{\no}{\nonumber}

\newcommand\lsim{\mathrel{\rlap{\lower4pt\hbox{\hskip1pt$\sim$}}
    \raise1pt\hbox{$<$}}}
\newcommand\gsim{\mathrel{\rlap{\lower4pt\hbox{\hskip1pt$\sim$}}
    \raise1pt\hbox{$>$}}}

\newcommand{\lra}{\longrightarrow}

\newcommand{\cL}{{\cal L}}
\newcommand{\cM}{{\cal M}}

\newcommand{\nn}{\nonumber \\}

\def\slch#1{\setbox0=\hbox{$#1$}\dimen0=\wd0%
\setbox1=\hbox{/}\dimen1=\wd1%
\ifdim\dimen0>\dimen1%
\rlap{\hbox to
\dimen0{\hfil/\hfil}}#1\else                                     
\rlap{\hbox to \dimen1{\hfil$#1$\hfil}}/\fi}

\newcommand{\mrm}{\mathrm}

\newcommand{\as}{\accentset}

\begin{document}
\parskip=3pt plus 1pt

\begin{titlepage}
\vskip 1cm
\begin{flushright}
{\sf UWThPh-2013-20}
\end{flushright}

\setcounter{footnote}{0}
\renewcommand{\thefootnote}{\fnsymbol{footnote}}

\vspace*{1.5cm}

\begin{center}
{\Large\bf Facets of chiral perturbation theory\footnote{To appear in
    the Proceedings of Hadron Structure '13, July 2013, Tatranske
    Matliare, Slovakia}
}
\\[20mm]

{\normalsize\bf Gerhard Ecker}\\[.6cm] 
 University of Vienna, Faculty of Physics \\ 
Boltzmanngasse 5, A-1090 Wien, Austria 
\end{center}

\vspace*{2cm}

\begin{abstract}
\noindent 
Chiral perturbation theory is the effective field theory of the
Standard Model at low energies. After a short introduction and
overview, I discuss three topics where the chiral approach leads to a
deeper understanding of low-energy hadron physics: radiative kaon
decays, carbogenesis in stellar nucleosynthesis and the interplay of
chiral perturbation theory and lattice QCD.
\end{abstract}

\setcounter{footnote}{0}
\renewcommand{\thefootnote}{\arabic{footnote}} 

\vfill

\end{titlepage}

\section{Motivation and overview}
\label{sec:intro}
For a systematic and quantitative treatment of the Standard Model (SM)
at low energies ($E < 1$ GeV), two approaches have survived the
scrutiny of time:
\begin{itemize} 
\item Effective Field Theory (EFT)
\item Lattice Field Theory
\end{itemize} 
The main objectives are to understand the physics of the SM in the
hadron sector at low
energies and to look for evidence of new physics.  
  
The low-energy region is not accessible in standard perturbation
theory because it is the strong-coupling regime of QCD. The key
concept for the EFT approach is the approximate chiral symmetry of
QCD:
\begin{equation}
\cL_{\rm QCD} = 
- \displaystyle\frac{1}{2} {\rm tr}
(G_{\mu\nu} G^{\mu\nu}) + \displaystyle\sum_{f=1}^{6}
\overline{q}_f \left(i \gamma^\mu D_\mu - m_f \mathbbm{1}_c \right)
q_f ~.
\end{equation}  
For massless quarks ($m_f = 0$), the chiral components
$q_{fL},q_{fR}$ can be rotated independently, leading to the chiral
symmetry of QCD with $n_F$ massless quarks  $SU(n_F)_L \times
SU(n_F)_R \times U(1)_V$.

Although $m_f = 0$ is a very good approximation for $n_F=2$ ($u,d$
quarks) and a reasonable one for $n_F=3$ ($u,d,s$),
there is no sign of chiral symmetry in the hadron spectrum. There are
many additional arguments pointing to the spontaneous breakdown of
chiral symmetry,
\begin{equation} 
SU(n_F)_L  \times SU(n_F)_R \times U(1)_V \lra
   SU(n_F)_{V} \times U(1)_V  ~,
\end{equation}
where the diagonal subgroup $SU(n_F)_{V}$ is either isospin ($n_F=2$)
or flavour $SU(3)$ ($n_F=3$). As a consequence, the spectrum of the
theory contains {$n_{F}^{2} - 1$ massless  Goldstone
  bosons. The associated fields parametrize the coset space
  $SU(n_F)_L  \times SU(n_F)_R~/~SU(n_F)_{V}$:
\begin{center} 
\begin{tabular}{ccccl}
 $n_{F}$ & \mbox{} \hspace*{.2cm} & $n_{F}^{2}-1$ &
\mbox{} \hspace*{.2cm} & Goldstone bosons \\
\hline
 2 & & 3 & & $\pi$ \\
 3 & & 8 & & $\pi, K, \eta$ \\
\hline \\
\end{tabular}
\end{center} 

Even in the real world with nonvanishing quark masses, pseudoscalar
meson exchange dominates amplitudes at low energies. For an EFT of
pseudo-Goldstone bosons only, chiral symmetry is realized nonlinearly
and the associated effective Lagrangian is necessarily
nonpolynomial. The EFT of the SM at low energies is called Chiral
Perturbation Theory (CHPT)
\cite{Weinberg:1978kz,Gasser:1983yg,Gasser:1984gg} and it is a
nonrenormalizable quantum field theory. Nevertheless, CHPT is a fully
renormalized QFT (in practice up to NNLO) and therefore independent of
the regularization procedure.

Another important consequence of Goldstone's theorem is at the basis
of the systematic low-energy expansion of CHPT: pseudo-Goldstone
bosons decouple for vanishing meson momenta and masses. The systematic
CHPT approach for low-energy hadron physics (for reviews, see
Refs.~\cite{Ecker:1994gg,Bernard:1995dp,Pich:1998xt,Scherer:2002tk,Bijnens:2006zp})
is  
\begin{itemize} 
\item most advanced in the meson sector (up to two loops, Table
  \ref{tab:Lagmeson}); 
\item it is also well developed for single-baryon and few-nucleon
  systems;
\item electroweak interactions can be and have been included.
\end{itemize}

\begin{table}
$$
\begin{tabular}{|l|c|} 
\hline
&  \\[-.2cm] 
\hspace{1cm} ${\cal L}_{\rm chiral\; order}$ 
~($\#$ of LECs)  &  loop  ~order \\[8pt] 
\hline 
&  \\
 ${\cal L}_{p^2}(2)$~+~ 
${\cal L}_{p^4}^{\rm odd}(0) $
~+~ ${\cal L}_{G_Fp^2}^{\Delta S=1}(2)$  
~+~ ${\cal L}_{G_8e^2p^0}^{\rm emweak}(1) $ & $L=0$
\\[3pt] 
~+~ ${\cal L}_{e^2p^0}^{\rm em}(1)$ ~+~
${\cal L}_{\mrm{kin}}^{\rm leptons}(0)$ & \\[10pt]
~+~ ${\cal L}_{p^4}(10)$~+~
${\cal L}_{p^6}^{\rm odd}(23)$
~+~${\cal L}_{G_8p^4}^{\Delta S=1}(22)$
~+~${\cal L}_{G_{27}p^4}^{\Delta S=1}(28)$ &   
$L \le 1$ \\[3pt]
~+~${\cal L}_{G_8e^2p^2}^{\rm emweak}(14)$ 
~+~ ${\cal L}_{e^2p^2}^{\rm em}(13)$
~+~ ${\cal L}_{e^2p^2}^{\rm leptons}(5)$  & \\[10pt] 
~+~ ${\cal L}_{p^6}(90)$  & $L \le 2$ \\[8pt] 
\hline
\end{tabular}
$$
\label{tab:Lagmeson}
\caption{Effective chiral Lagrangian in the meson sector for chiral
  $SU(3)$. In brackets, the number of coupling constants (LECs) of
  CHPT.} 
\end{table}

For this talk, I have chosen three topics where the main emphasis is on 
obtaining a better understanding of hadronic interactions at low
energies rather than on high-precision studies with the potential to
look for evidence of new physics (e.g., in semileptonic kaon
decays). Theoretical and experimental
investigations of the radiative kaon decays $K_S \to \gamma\gamma$ and
$K_L \to \pi^0 \gamma \gamma$  span a period of more than a quarter
century, from the second half of the 80s of last century where
CHPT was just one 
of many ``hadronic models'' to fairly recent times where CHPT predictions
have been verified experimentally. An interesting application of chiral
EFTs in nuclear physics is the recent attempt to quantify the
sensitivity of the so-called Hoyle state to fundamental
parameters of the SM, the light quark mass and the electromagnetic
fine-structure constant. The results add a new touch to the
understanding of the abundance of carbon and oxygen in the universe in
terms of the anthropic principle. Finally, to illustrate the fruitful
collaboration between the two main players in low-energy hadron
physics, CHPT and lattice QCD, I discuss ongoing attempts to extract
information on some low-energy constants (LECs) from lattice simulations. I
present preliminary results of an approach making use of an analytic
approximation of two-loop amplitudes in chiral $SU(3)$.

\vspace*{.2cm} 
\section{Nonleptonic kaon decays}
\label{sec:kaon}
Kaon decays are a fertile field for CHPT (for a general review, see
Ref.~\cite{Cirigliano:2011ny}) . While in some semileptonic decays the
precision provided by CHPT allows to search for evidence of new
physics, the situation is much more complicated in nonleptonic
decays. Nevertheless, a comprehensive picture has emerged over the
years through the collaboration between theory and experiment. In this
section, I briefly review the status of a specific subclass of
radiative kaon decays.

The two basic couplings of the leading-order nonleptonic chiral
Lagrangian ${\cal L}_{G_Fp^2}^{\Delta S=1}(2)$, usually called $G_8,
G_{27}$, are by now well established from studies of the dominant
nonleptonic kaon decays $K \to 2 \pi, 3 \pi$ up to NLO, including
isospin-violating and radiative corrections
\cite{Cronin:1967jq,Kambor:1991ah,Cirigliano:2003gt,Bijnens:2004ku,Bijnens:2004vz,Bijnens:2004ai}. 
\begin{center} 
\begin{figure}[!ht]
\begin{center}  
\leavevmode 
\includegraphics[width=9cm]{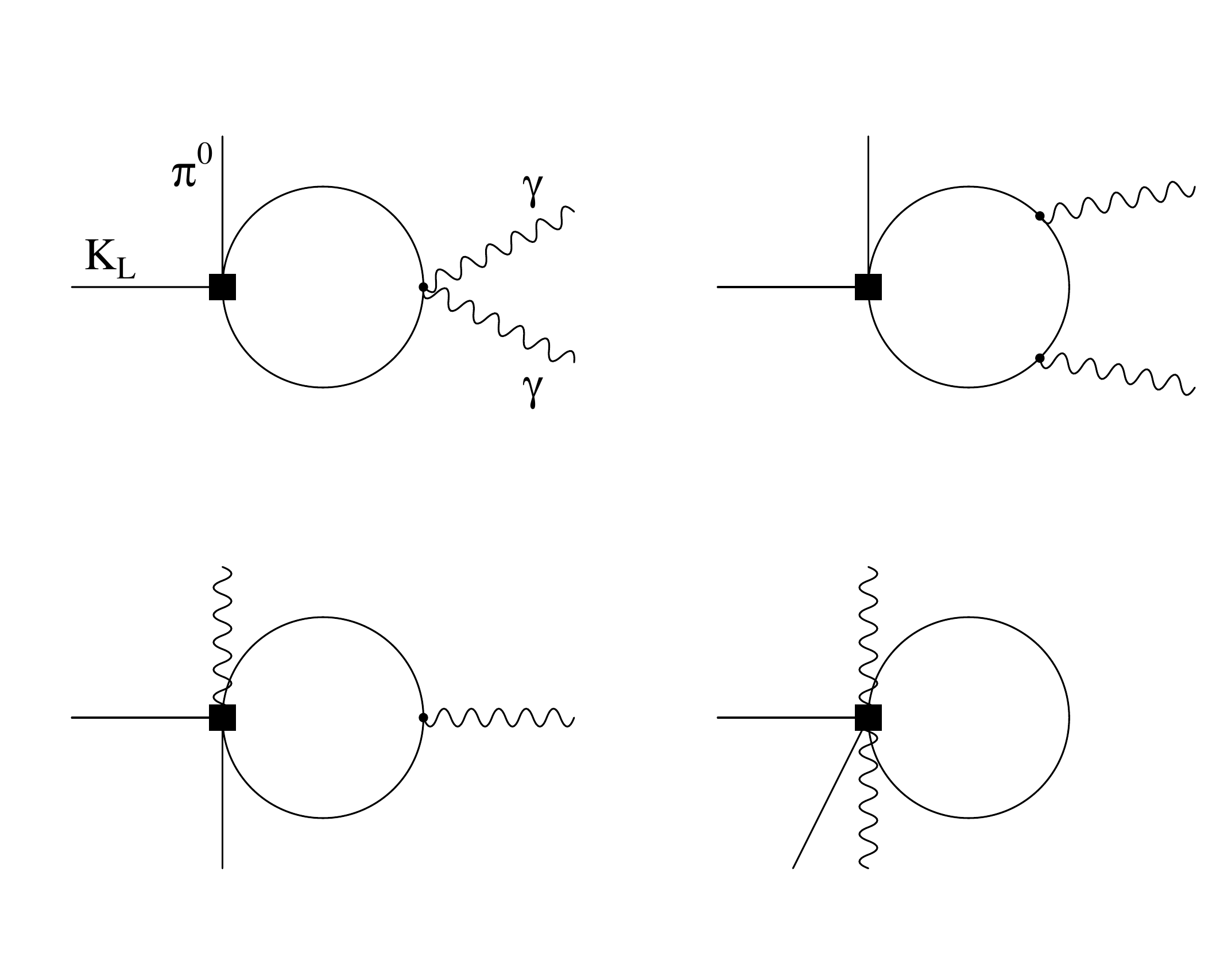}
\end{center}
\caption{One-loop diagrams for $K_L \to \pi^0 \gamma \gamma$
\cite{Ecker:1987fm,Cappiello:1988yg}. For $K_S
\to \gamma\gamma$, replace $K_L$ by $K_S$ and drop the $\pi^0$ line
\cite{D'Ambrosio:1986ze,Goity:1986sr}.}
\label{fig:klpigg}
\end{figure}
\end{center}
All other nonleptonic kaon decays start at NLO, $O(G_F p^4)$, only. As
indicated in Table \ref{tab:Lagmeson}, there are 22 (octet) plus 28
(27-plet) new LECs entering at NLO. Therefore, the radiative decays
$K_S \to \gamma\gamma$,  $K_L \to \pi^0 \gamma \gamma$ and  $K_S \to
\pi^0 \pi^0 \gamma \gamma$ have been especially popular among
CHPT theorists: none of the $22 ~+~ 28$ NLO LECs contributes! 
Therefore, at NLO the decay amplitudes are given by finite one-loop
contributions in terms of the known LO couplings $G_8,G_{27}$ only.
The channel  $K_L \to \pi^0 \gamma \gamma$ is not only interesting in
its own right because it generates a CP-conserving contribution via the
two-photon cut to the dominantly CP-violating decays $K_L \to \pi^0
l^+ l^-$ \cite{Donoghue:1987awa,Ecker:1987hd,Sehgal:1988ej}.

In the remainder of this section, I review the status of the decays
$K_S \to \gamma\gamma$ and  $K_L \to \pi^0 \gamma \gamma$ (the decay $K_S \to
\pi^0 \pi^0 \gamma \gamma$ \cite{Funck:1992wa} has not been observed
yet). At $O(G_F p^4)$, the relevant diagrams are shown in
Fig.~\ref{fig:klpigg}. Note that each one of the diagrams is
quadratically divergent: chiral symmetry ensures that the sum is finite.  

As predicted by CHPT, already the first observation of $K_L \to \pi^0 \gamma
\gamma$ \cite{Barr:1990hc} demonstrated that the two-photon spectrum
is dominated by the pion-loop 
contribution, in contrast to the previously assumed vector meson
dominance. However, it also became clear that the rate was
underestimated. Higher-order corrections needed to be taken into
account, starting at $O(G_F p^6)$.
\begin{center} 
\begin{figure}[!h] 
\begin{center} 
\includegraphics[width=9cm]{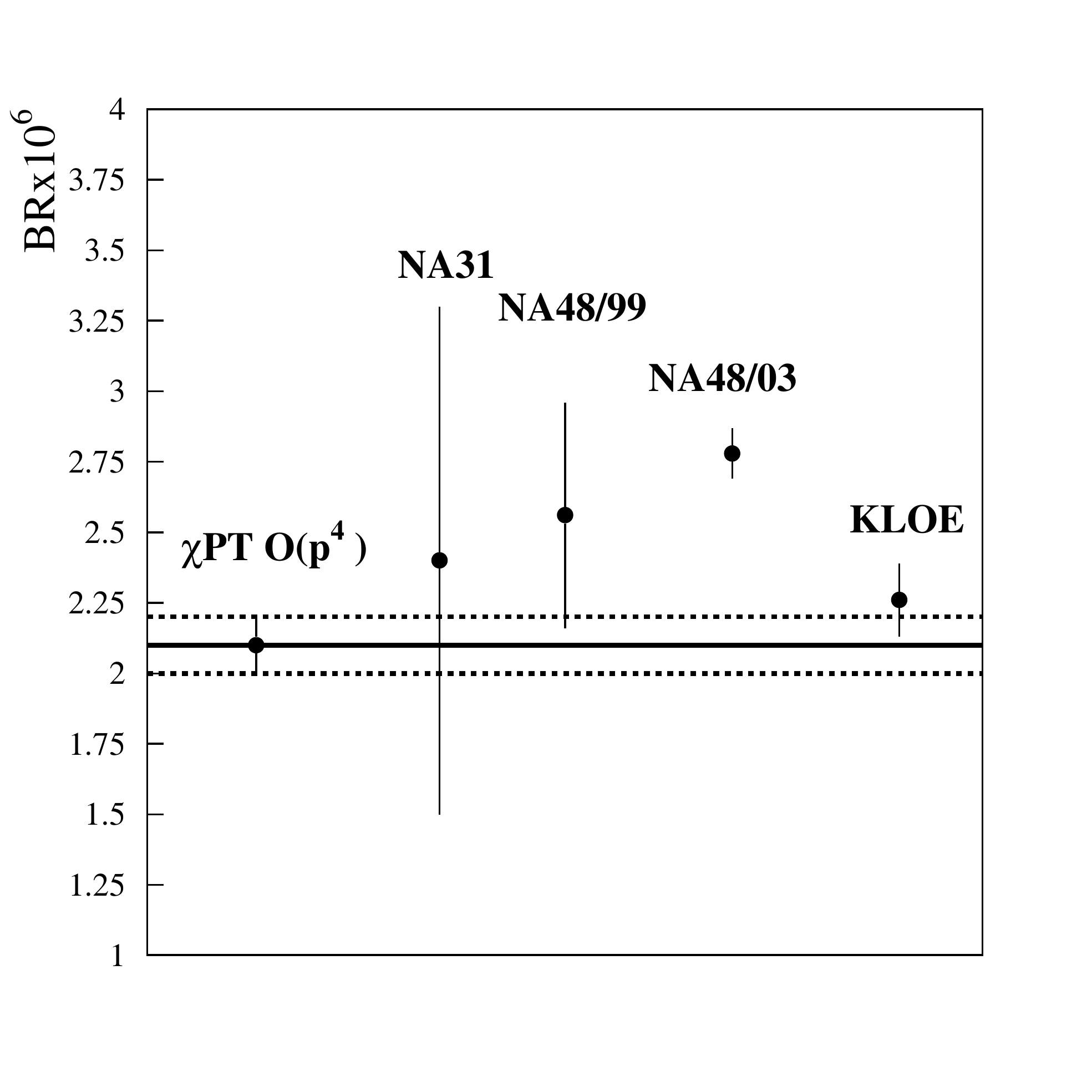}
\end{center}
\caption{Experimental results for the branching ratio $B(K_S \to
  \gamma \gamma)$ in comparison with the chiral prediction.} 
\label{fig:ksgg}
\end{figure}
\end{center} 
\begin{enumerate} 
\item[i.] Rescattering (unitarity) corrections turned out to be small
  for $K_S \to \gamma\gamma$ \cite{Kambor:1993tv}, but they are sizable in
  the case of $K_L \to \pi^0 \gamma \gamma$
  \cite{Cappiello:1992kk,Cohen:1993ta}.
\item[ii.] Resonance contributions were estimated to be small for
  the $K_S$ decay mainly because vector mesons cannot contribute. This
  is again different for the $K_L$ decay: although the vector meson
  contribution is model dependent, it is to a good approximation
  parametrized by a single parameter $a_V$
  \cite{Cohen:1993ta,D'Ambrosio:1996sw}. 
\end{enumerate} 
\begin{center} 
\begin{figure}[!h] 
\begin{center} 
\leavevmode 
\includegraphics[width=9cm]{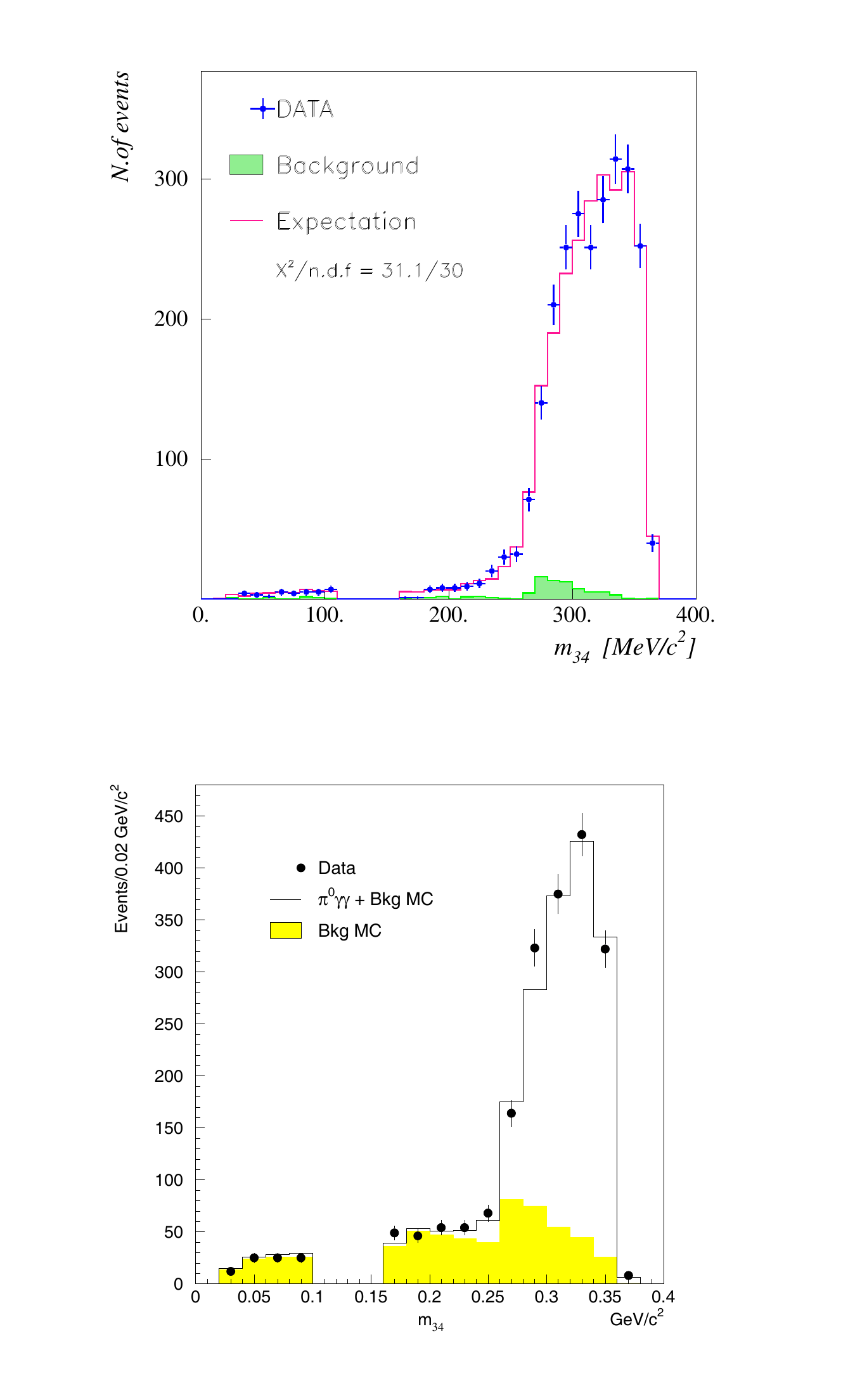}
\end{center}
\caption{Two-photon spectra for $K_L \to \pi^0 \gamma \gamma$ from
  NA48 \cite{Lai:2002kf} (top) and KTeV \cite{Abouzaid:2008xm} (bottom).}
\label{fig:na48ktev} 
\end{figure}
\end{center} 
It therefore came as a surprise when NA48 \cite{Lai:2002sr} announced
a rate for  
$K_S \to \gamma\gamma$ substantially bigger than the chiral prediction
(see Fig.~\ref{fig:ksgg}). Fortunately, the more recent result of KLOE
\cite{Ambrosino:2007ab}, $B(K_S \to \gamma \gamma) = 2.26(12)(06) \times
10^{-6}$, is again in perfect agreement with expectations. The
decision by the Particle Data Group \cite{pdg:2012} to average the
results of NA48 and KLOE does not appear very illuminating: another
experiment is needed to clarify the issue.

After several years of discrepancies, the experimental situation for
$K_L \to \pi^0 \gamma \gamma$ has now been clarified
\cite{Lai:2002kf,Abouzaid:2008xm}. Both the two-photon spectra shown
in Fig.~\ref{fig:na48ktev} and the branching ratios agree among each other
and with CHPT \cite{pdg:2012}:
\begin{eqnarray} 
B(K_L \to \pi^0 \gamma \gamma) \cdot 10^6 &=& 
1.273 \pm 0.033  \no \\[.1cm]
a_V &=& -0.43 \pm 0.06  ~.
\end{eqnarray}
As an important by-product of this result, the CP-conserving
contribution $K_L \to \pi^0 \gamma^* \gamma^* \to \pi^0 e^+ e^-$ is
indeed negligible in comparison with the CP-violating amplitudes.

\section{Carbogenesis}
\label{sec:hoyle}
\begin{center} 
\begin{figure}[!h]
\begin{center}  
\leavevmode
\includegraphics[width=9cm]{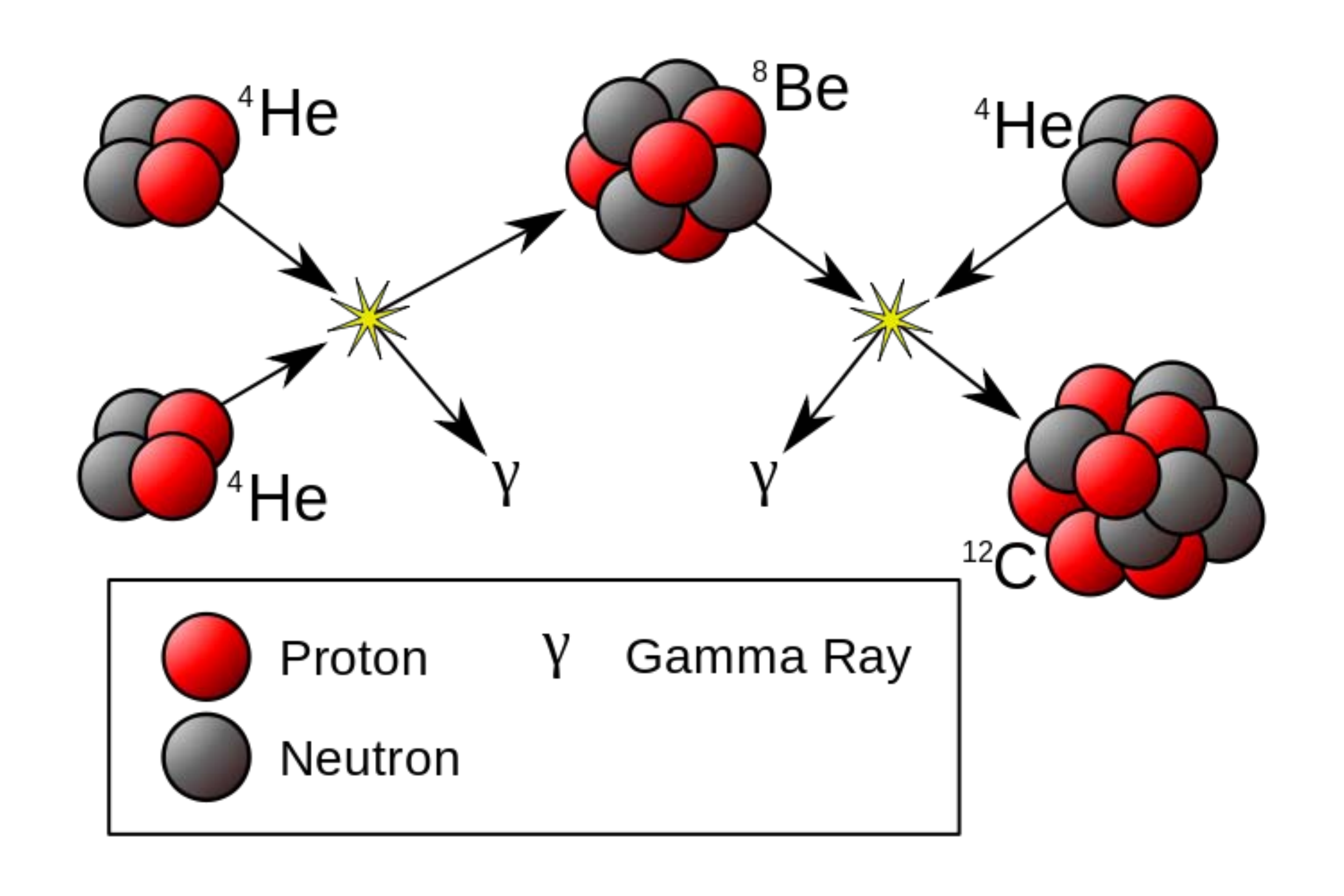} 
\end{center} 
\caption{Triple-$\alpha$ process for $^{12}$C.}
\label{fig:hoyle}
\end{figure}
\end{center}
Almost all carbon in the universe is produced in stellar
nucleosynthesis via the triple-$\alpha$ process shown in
Fig.~\ref{fig:hoyle}. 
In order to explain the observed carbon abundance, Hoyle
\cite{Hoyle:1954zz} postulated the existence of an excited state of
$^{12}$C near the $^{8}$Be-$\alpha$ threshold that was observed soon
afterwards. Two important characteristics of this resonance are the
energy $\epsilon$ above the $3\alpha$ threshold and its radiative
width $\Gamma_\gamma$:
\begin{eqnarray} 
\epsilon=379.47(18) ~{\rm keV}, \quad & \Gamma_\gamma=3.7(5)
~{\rm meV}~.  \label{eq:hoyle}
\end{eqnarray} 
Since the triple-$\alpha$ rate is proportional to $\Gamma_\gamma
\exp{-\epsilon/kT}$, the rate is most sensitive to $\epsilon$. This
sensitivity has often been considered a prime example of the anthropic
principle, but later investigations showed that a difference $\Delta
\epsilon \lsim 100$ keV could be tolerated to explain the abundance of
$^{12}$C and $^{16}$O \cite{Livio:1989,Oberhummer:2000zj}.

Although this range cannot be considered extreme fine-tuning, the more
interesting issue is the dependence of $\epsilon$ on fundamental
parameters of strong and electromagnetic interactions. Using a
one-parameter nuclear cluster model, Schlattl et
al. \cite{Schlattl:2003dy} found that the tolerances for the strength
parameter $p$ and the Coulomb force $F_{\rm Coulomb}$ were indeed
small: 
\begin{eqnarray} 
\Delta p/p \lsim 0.5 \%, \quad &  \Delta F_{\rm Coulomb}/F_{\rm
    Coulomb} \lsim 4 \% ~.
\label{eq:oberh}
\end{eqnarray}    
However, the quantities $p$ and $F_{\rm Coulomb}$ in the model of
Ref.~\cite{Schlattl:2003dy} are difficult to relate
to fundamental parameters of QCD and QED. CHPT can provide at least a
partial solution of this problem. 

The chiral EFT of nuclear forces put forward by Weinberg
\cite{Weinberg:1990rz} has proven to 
be very successful for small nuclei ($A \lsim 3$). 
In a more recent development, the nuclear chiral EFT has been put on
the lattice (for a review of nuclear lattice simulations, see
Ref.~\cite{Lee:2008fa}). The important difference to lattice QCD (see
Sec.~\ref{sec:lattice}) is that instead of quarks and gluons the
lattice degrees of freedom are now nucleons and pions. The nuclear
simulations have been quite successful in calculating energy spectra
of light nuclei. As an example, I reproduce in Table
\ref{tab:spectrum} the low-lying even-parity spectrum of $^{12}$C
calculated by Epelbaum et al. \cite{epelbaum:2012qn}.    
\begin{center} 
\begin{table}[!h]
\begin{center} 
\begin{tabular}{|c||c|c|c|} \hline
& \hspace{.2cm} $0_{1}^{+}$ \hspace{.2cm}
&  $2_{1}^{+}(E^{+})$  
& \hspace{.2cm} $\textcolor{red}{0_{2}^{+}}$ 
 \\ \hline
LO & $-96(2)$ & $-94(2)$ & $\textcolor{red}{-89(2)}$  \\
NLO & $-77(3)$ & $-74(3)$ & $\textcolor{red}{-72(3)}$  \\
NNLO & $-92(3)$ & $-89(3)$ & $\textcolor{red}{-85(3)}$  \\ \hline
Exp & $-92.16$ & $-87.72$ & $\textcolor{red}{-84.51}$ 
\\ \hline
\end{tabular}
\end{center}
\caption{Energies of low-lying even-parity states of $^{12}$C (in MeV)
  from Ref.~\cite{epelbaum:2012qn}. $0_{2}^{+}$ is the Hoyle resonance.} 
\label{tab:spectrum}
\end{table}
\end{center} 
With nuclear CHPT one cannot study the influence of the strong
coupling $\alpha_{\rm QCD}$ (hidden in nucleons and pions), but the
impact of the light quark mass $m_q$ in the isospin limit (via
$M_\pi^2 \sim (m_u+m_d)$ at lowest order CHPT) and of
the fine-structure constant $\alpha_{\rm em}$ can be investigated. The
final conclusion obtained in Ref.~\cite{Epelbaum:2012iu} is that the
necessary fine-tuning of $m_q$ and $\alpha_{\rm em}$ is much more severe
than for the energy difference $\epsilon$ in Eq.~(\ref{eq:hoyle}): 
\begin{eqnarray}
\Delta m_q/m_q \lsim 3 \%~, \quad & \Delta \alpha_{\rm em}/\alpha_{\rm
   em} \lsim 2.5 \%~.
\end{eqnarray}   
While the constraint on the fine-structure constant is in accordance
with the previous result in Eq.~(\ref{eq:oberh}), the allowed range for
the light quark mass adds another touch to the interpretation of the
anthropic principle.

\section{Low-energy constants and lattice QCD}
\label{sec:lattice}
In recent years, the collaboration between the two major
players in low-energy hadron physics, CHPT and lattice QCD, has 
intensified considerably.
\begin{itemize} 
\item[i.] Extrapolation to the physical quark (and meson) masses
  provided by CHPT is still useful for lattice simulations, but
  because of more powerful computers less so than some five years
  ago. On the other hand, finite-volume corrections accessible in CHPT
  are still needed for a reliable estimate of lattice
  uncertainties. 
\item[ii.] On the other hand, the determination of LECs from lattice
  studies has become more important over the years. This input is
  especially welcome for those LECs that modulate quark mass terms:
  unlike in standard phenomenological analysis, the lattice physicist
  can tune quark (and therefore meson) masses.
\end{itemize}    
The present situation can be characterized by the following motto
\cite{LL:2013}, modeled after a famous quote: ``Ask not what
CHPT can do for the lattice, but ask what the lattice can do for CHPT''.

As an illustrative example, consider one of the two
leading-order LECs, the meson decay constant in the chiral limit. The
chiral $SU(2)$ LEC $F =  \lim_{m_u,m_d \to 0} F_\pi$ is well
known, mainly from a combined analysis of lattice data by the FLAG
Collaboration \cite{Colangelo:2010et,FLAG:2013}: 
\begin{equation} 
F = (85.9 \pm 0.6) ~{\rm MeV}~.
\label{eq:F}
\end{equation}
The situation is different in the $SU(3)$ case. The lattice
results for $F_0 =  \lim_{m_u,m_d,m_s \to 0} F_\pi$ cover a much wider
range, from about 66 MeV to 84 MeV \cite{FLAG:2013}. Consequently, the
FLAG group refrains from performing an average. A similar 
range is covered in the phenomenological fits of Bijnens and
Jemos \cite{Bijnens:2011tb} as shown in Fig.~\ref{fig:bj}. 
\begin{center} 
\begin{figure}[!ht]
\begin{center}  
\leavevmode
\includegraphics[width=9cm]{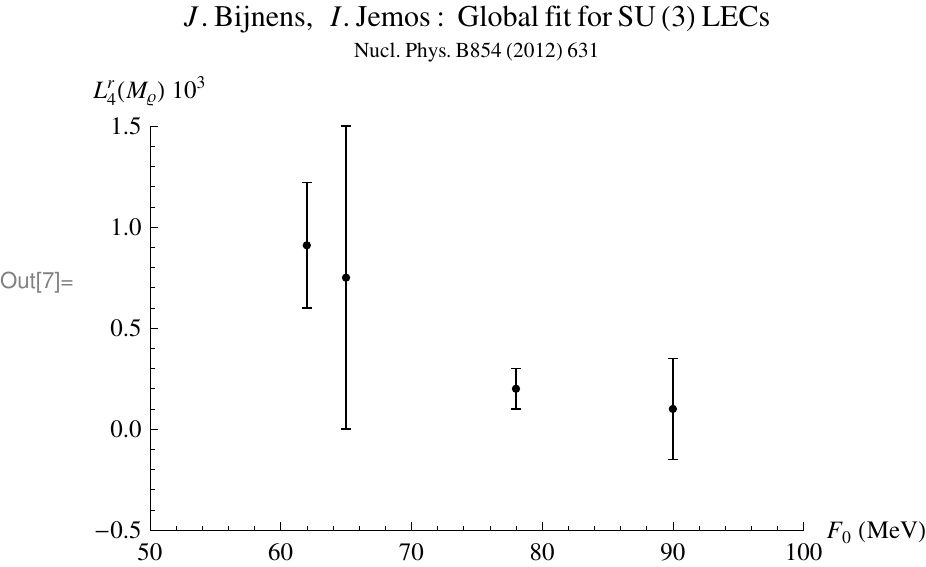} 
\end{center} 
\caption{Results for $L_4^r(M_\rho)$ and $F_0$ from various fits in
  Ref.~\cite{Bijnens:2011tb}. $F_0$ is not fitted but calculated a
  posteriori from the fitted LECs.}
\label{fig:bj}
\end{figure}
\end{center}
The low-energy expansion in chiral $SU(3)$ is characterized by the
ratio $p^2/(4\pi F_0)^2$ where $p$ stands for a generic meson momentum
or mass. The LEC 
$F_0$ thus sets the scale for the chiral expansion. In practical work,
$F_0$ is usually traded for $F_\pi$ at successive orders of the chiral
expansion. Nevertheless, $F_0$ sets the scale of ``convergence'' of
the chiral expansion: a smaller $F_0$ tends to produce bigger
fluctuations at higher orders. It is therefore disturbing that its
value is less known than for many higher-order LECs.

One clue to the difficulty of extracting $F_0$ is the apparent
anti-correlation with the NLO LEC $L_4$ in the fits of
Ref.~\cite{Bijnens:2011tb}: the bigger $F_0$, the smaller
$L_4^r(M_\rho)$, and vice versa (see Fig.~\ref{fig:bj}). The
large-$N_c$ suppression of $L_4$ is not manifest in the fits with
small $F_0$. 

This anti-correlation can be understood to some extent from the
structure of the chiral $SU(3)$ Lagrangian up to and including NLO
(see Table \ref{tab:Lagmeson}):  
\noindent
\begin{eqnarray}
{\cal L}_{p^2}(2) + {\cal L}_{p^4}(10) &=&
\displaystyle\frac{F_0^2}{4} \langle 
D_\mu U D^\mu U^\dagger +  \chi U^\dagger + \chi^\dagger  U \rangle 
+ L_4 \langle D_\mu U D^\mu U^\dagger\rangle 
\langle \chi U^\dagger + \chi^\dagger  U \rangle + \dots  \nn
&=& \displaystyle\frac{1}{4} \langle D_\mu U
D^\mu U^\dagger \rangle \left[ F_0^2 + 8 L_4 \left(2 \as{\circ}{M}^2_K +
  \as{\circ}{M}^2_{\pi} \right)\right] + \dots 
\end{eqnarray} 
where $U = \mathds{1} ~+$ ~meson fields, $\chi = 2B_0 \cM_q$ ($B_0
\sim$ quark condensate, $\cM_q$ is the quark mass matrix), $\langle \dots
\rangle$ stands for the $SU(3)$ flavour trace and 
$\as{\circ}{M}_P$ denotes the lowest-order meson masses. The dots 
refer to the remainder of the NLO Lagrangian in the first
line and to terms of higher order in the meson fields in the second
line. Therefore, a LO tree-level contribution is always
accompanied by an $L_4$ contribution in the combination 
\begin{equation} 
F(\mu)^2:=F_0^2 + 8 L_4^r(\mu) \left(2 \as{\circ}{M}^2_K +
\as{\circ}{M}^2_{\pi}\right)~.
\label{eq:Fmu}
\end{equation} 
Of course, there will in general be additional contributions involving
$L_4$ at NLO, especially in higher-point functions (e.g., in meson
meson scattering). Nevertheless, the observed anti-correlation between
$F_0$ and $L_4$ is clearly related to the structure of the chiral
Lagrangian. Note that $F^2_{\pi}/(16 M_K^2) = 2 \times 10^{-3}$ is the
typical size of a NLO LEC. Although of different chiral order, the two
terms in $F(\mu)^2$ could a priori be of the same order of magnitude.
   
Independent information on $F_0$ comes from comparing the $SU(2)$ and
$SU(3)$ expressions for $F_\pi$.  To $O(p^4)$ in chiral $SU(2)$,
$F_\pi$ is given by \cite{Gasser:1983yg} 
\begin{equation} 
F_\pi = F + F^{-1} \left[M_\pi^2 \,l_4^r(\mu) + \overline{A}(M_\pi,\mu)
\right]
\label{eq:su2}
\end{equation}
where $l_4$ is a NLO $SU(2)$ LEC and $\overline{A}(M_\pi,\mu)$ is a
one-loop function.
Expressing $l_4^r(\mu)$  in terms of $L_4^r(\mu)$, $L_5^r(\mu)$
and a kaon loop contribution \cite{Gasser:1984gg} and equating
Eq.~(\ref{eq:su2}) with the $SU(3)$ result for $F_\pi$
\cite{Gasser:1984gg}, one arrives at the following relation:
\begin{eqnarray}  
F_0 &=& F - F^{-1}\left\{\left(2 M_K^2 - M_\pi^2\right)\left(4 L_4^r(\mu) +
\displaystyle\frac{1}{64 \pi^2}\log{\displaystyle\frac{\mu^2}{M_K^2}}\right)
+ \displaystyle\frac{M_\pi^2}{64 \pi^2}  \right\} + O(p^6) ~.
\label{eq:F0F}
\end{eqnarray} 
Assuming the ``paramagnetic'' inequality $F_0 < F$
\cite{DescotesGenon:1999uh} to hold already at $O(p^4)$, one gets a lower
bound for $L_4$, 
\begin{equation} 
 L_4^r(M_\rho) > - 0.4 \times 10^{-3}~,
\end{equation} 
well compatible with existing estimates.

$SU(3)$ lattice data for $F_\pi$ seem well suited for a determination
of $F_0$ and $L_4$. For a quantitative analysis, the use of CHPT to
NNLO, $O(p^6)$, is essential. In many analyses of lattice data, the
complete NNLO result for $F_\pi$ in chiral $SU(3)$
\cite{Amoros:1999dp}, which is available in numerical form only, has not
been employed so far.
Some time ago, we proposed a large-$N_c$ motivated approximation for
NNLO calculations in chiral $SU(3)$ where the loop amplitudes are
given in analytic form \cite{Ecker:2010nc}. In the remainder of this
section, I report on a preliminary analysis of $F_\pi$ within
this framework to extract the LECs  $F_0$, $L_4$ \cite{emn:2013}.

The following input is needed. The two-loop contribution depends on a
single additional parameter $M$, the scale of double logs. Comparing
with a numerical analysis \cite{Bernard:2009ds}, we have
convinced ourselves that $M \simeq M_K$ as expected, at least for
$F_K/F_\pi$ and for $F_\pi$ 
itself. In addition, some knowledge of the only other LEC $L_5$
appearing at NLO and of the other LECs entering at $O(p^6)$ is
required. The following (preliminary) results \cite{emn:2013} take the
uncertainties of $M$ and the LECs involved into account, adding
errors in quadrature to the lattice errors. We use lattice data for
$F_\pi$ from the RBC/UKQCD Collaboration
\cite{Aoki:2010dy,Arthur:2012opa}. 
\begin{center} 
\begin{figure}[!ht]
\begin{center}  
\leavevmode
\includegraphics[width=10cm]{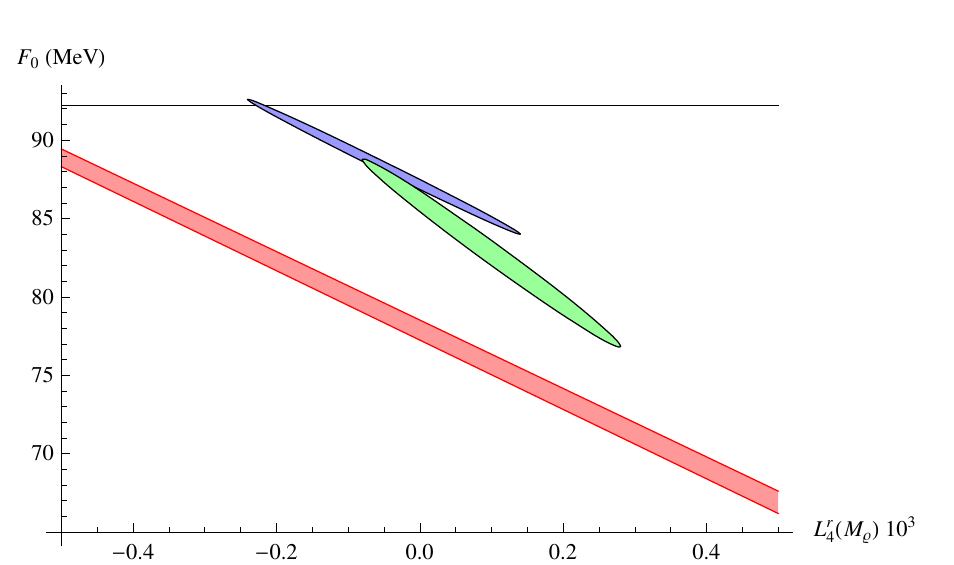} 
\end{center} 
\caption{Fitted values of $F_0$, $L_4$ using RBC/UKQCD data
  \cite{Aoki:2010dy,Arthur:2012opa} with $M_\pi < 350$ MeV, with (blue
  ellipse) and without (green ellipse) including $F_\pi^{\rm
  phys}$. The red band results from the  
  comparison of $F_\pi$ between $SU(2)$ and $SU(3)$ as expressed by
  Eq.~(\ref{eq:F0F}), taking $F  = (85.9 \pm 0.6) ~{\rm MeV}$ from the
  FLAG compilation \cite{FLAG:2013}. The horizontal line denotes
  $F_\pi = 92.2$ MeV.}
\label{fig:F0FL4_34}
\end{figure}
\end{center}
The extracted values of $F_0$ and $L_4^r(M_\rho)$ are shown in
Fig.~\ref{fig:F0FL4_34}. The ellipses describe two options, depending
on whether the physical value of $F_\pi$ is included in the fit (blue
ellipse) or not (green ellipse). The red band originates from the
comparison between chiral $SU(2)$ and $SU(3)$ as expressed by
Eq.~(\ref{eq:F0F}). Referring to Ref.~\cite{emn:2013} for a more
complete discussion, I list the values of $F_0$ and $L_4^r(M_\rho)$
corresponding to the blue ellipse ($F_\pi^{\rm phys}$ included):
\begin{eqnarray}
F_0 &=& (88.3 \pm 4.3) ~{\rm MeV} \nn
10^3 L_4^r(M_\rho) &=& - 0.05 \pm 0.19 \nn
{\rm corr}(F_0,L_4^r) &=& - 0.997 ~.
\label{fig:fit3}
\end{eqnarray} 
\begin{center} 
\begin{figure}[!hb]
\begin{center}  
\leavevmode
\includegraphics[width=9cm]{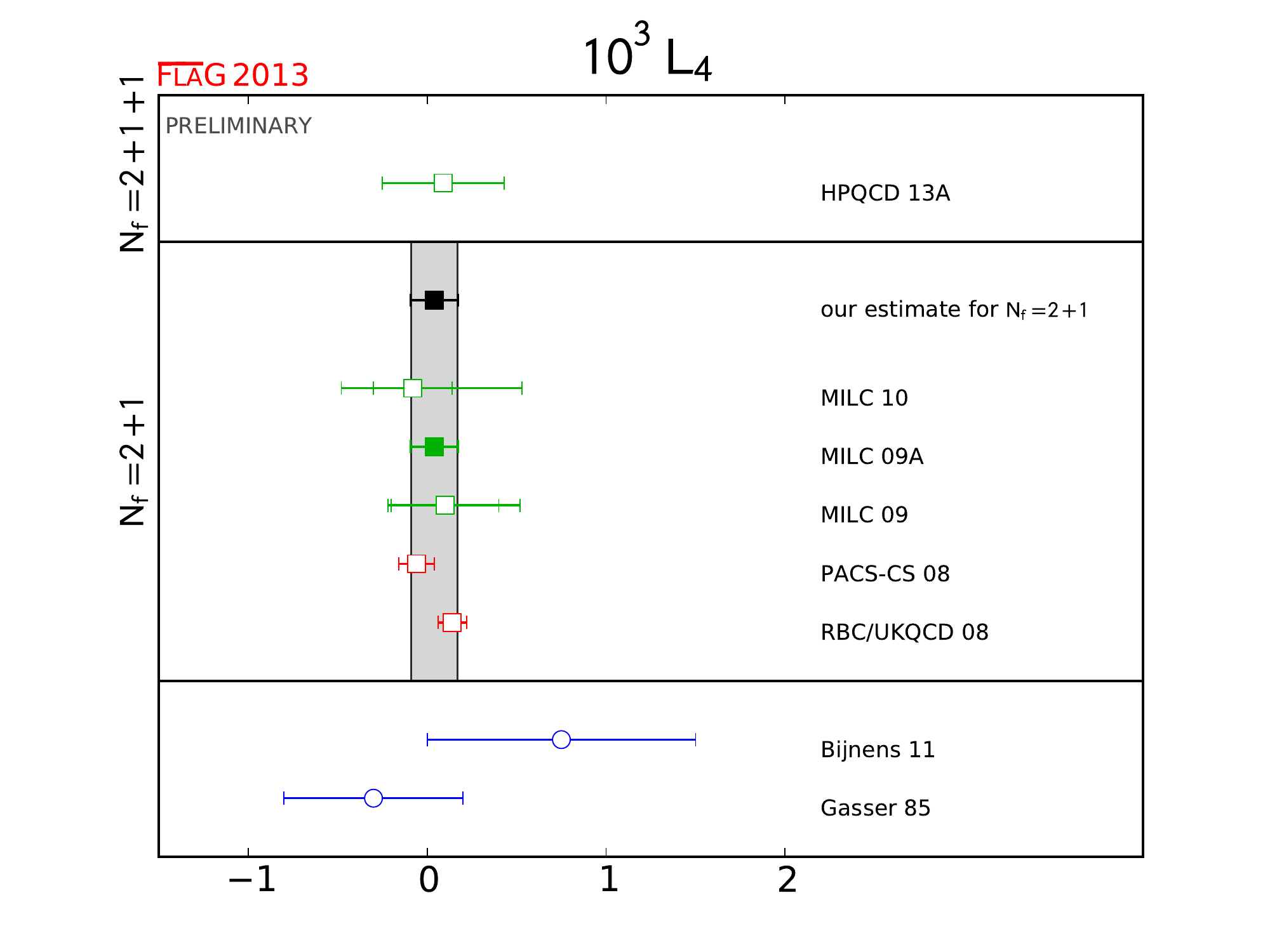} 
\end{center} 
\caption{A compilation of lattice determinations of 
  $L_4^r(M_\rho)$ \cite{FLAG:2013}.}
\label{fig:L4}
\end{figure}
\end{center} 
The two ellipses are roughly compatible with each other. The green
ellipse is a little 
lower because from the RBC/UKQCD data alone the fitted value of
$F_\pi$ is slightly smaller than the experimental value. The value
for $L_4$ is consistent with large $N_c$ and with available lattice
results \cite{FLAG:2013} shown in Fig.~\ref{fig:L4}. The result for
$F_0$ is more precise than both phenomenological
(cp. Fig.~\ref{fig:bj}) and existing lattice determinations
\cite{Colangelo:2010et,FLAG:2013}. It is a little bigger than expected
\cite{DescotesGenon:1999uh}, approximately of the same size as the
$SU(2)$ LEC $F$ in Eq.~(\ref{eq:F}). Moreover, $F_0$ in
Eq.~(\ref{fig:fit3}) does not match with the comparison
between $SU(2)$ and $SU(3)$ as indicated by the red band in
Fig.~\ref{fig:F0FL4_34}.  

Aside from possibly underestimated uncertainties, this discrepancy may
be due to the fact that the red band in Fig.~\ref{fig:F0FL4_34} is
based on $O(p^4)$ calculations whereas the fit values in
Eq.~(\ref{fig:fit3}) result from an (albeit approximate) calculation
to $O(p^6)$. Note also that the value for $F$ in Eq.~(\ref{eq:F}) is
an average 
over all existing lattice results; the most precise determinations
with $N_f=2+1$ active flavours produce a slightly bigger average $F =
(86.8 \pm 0.3)$ MeV \cite{FLAG:2013}. Nevertheless, the discrepancy
between the direct fit (\ref{fig:fit3}) and the $SU(2)$ constraint
(\ref{eq:F0F}) would remain.   

The strong anti-correlation between $F_0$ and $L_4$ persists because
the kaon masses in the RBC/UKQCD data are all close to the physical
kaon mass. Simulations with smaller kaon masses \cite{Bazavov:2009bb}
would not only be
welcome from the point of view of convergence of the chiral series,
but they could also 
provide a better lever arm for reducing the anti-correlation and the
fit errors of $F_0$ and $L_4$. This expectation is justified because
the quantity $F(M_\rho)$ defined in Eq.~(\ref{eq:Fmu}) is much
better determined than $F_0$.

\vspace*{.2cm} 
\section{Conclusions}
\label{sec:conclusions}
We started out with stating the main objectives of CHPT:
to understand the structure of the SM at low energies
and to look for possible evidence of new physics. Have these
objectives been accomplished?

We have certainly come some way with CHPT in understanding hadronic
interactions at low energies. I discussed three examples where the
CHPT approach has given rise to significant new insights. The
investigation of the radiative nonleptonic kaon decays $K_S \to
\gamma\gamma$ and  $K_L \to \pi^0 \gamma \gamma$ during the past 25
years has led to an overall agreement between theory and
experiment, with a minor discrepancy between two experiments for the
$K_S$ decay still pending. Nuclear physics for light nuclei has 
made impressive progress with the help of chiral EFTs. A recent
formulation on the lattice (with nucleons and pions) seems very
promising. As an example, the impact of the light quark masses and of the
fine-structure constant on the Hoyle resonance in $^{12}$C was studied
with such an approach. Finally, the interaction between CHPT and
lattice QCD is prospering. Many of the CHPT couplings that are
difficult to obtain from phenomenology are now becoming accessible on
the lattice.  

Concerning the second objective mentioned in the introduction, we have
not found any evidence for new physics with CHPT. But neither has the
LHC! 


\end{document}